\documentclass{emulateapj}
\usepackage{amsmath}
\usepackage{amssymb}
\usepackage{natbib}
\usepackage{graphicx}
\usepackage{color}
\usepackage {ulem}

\shorttitle{The Mystery of the $\sigma$-Bump}
\shortauthors{Schauer et al.}

\newcommand{\Reff}{R_{\mathrm{eff}}}

\begin{document}

\title{The Mystery of the $\sigma$-Bump -- A new Signature for Major Mergers in Early-type Galaxies?}

\author{Anna Therese Phoebe Schauer$^1$, Rhea-Silvia Remus$^1$$^2$, Andreas Burkert$^1$$^2$ and Peter H. Johansson$^3$}
\affil{$^1$ Universit\"ats-Sternwarte M\"unchen, Scheinerstr.\ 1, D-81679 M\"unchen, Germany\\
$^2$ Max-Planck-Institute for Extraterrestrial Physics, P.O. Box 1312, D-85748 Garching, Germany\\
$^3$ Department of Physics, University of Helsinki, Gustaf H\"allstr\"omin katu 2a, FI-00014 Helsinki, Finland\\
\texttt{aschauer@usm.uni-muenchen.de} \\}

\begin{abstract}
The stellar velocity dispersion as a function of the galactocentric radius of an early-type galaxy can generally be well approximated 
by a power law $\sigma  \propto  r^{\beta}$. 
However, some observed dispersion profiles show a deviation from this fit 
at intermediate radii, usually between one and three $\Reff$, 
where the velocity dispersion remains constant with radius, showing a bump-like behavior, 
which we term the ``$\sigma$-''bump. 
To understand the origin of this 
$\sigma$-bump, we study a set of simulated early-type galaxies formed in major mergers. 
We find the $\sigma$-bump in all of our simulated early-type galaxies, with the size and position of the
bump slightly varying from galaxy to galaxy, leading to the assumption that the bump is a 
characteristic of the major merger formation scenario.
The feature can be seen both in the intrinsic and projected stellar velocity dispersions. 
In contrast to shells that form during the merger event but evolve with time and finally disappear, 
the $\sigma$-bump stays nearly constant 
with radius and is a permanent feature that is preserved until the end of the simulation. 
The $\sigma$-bump is not seen in the dark matter and gas components and
we therefore conclude that it is a purely stellar feature of merger remnants. 
\end{abstract}


\section{Introduction}
The kinematics of early-type galaxies (ETGs) has been a subject of interest for many years. 
Unfortunately, the stellar light decreases rapidly after one to two effective radii, 
allowing observations to reliably detect only the inner effective radii 
\citep{proctor:2009MNRAS.398...91P,foster:2013arXiv1308.3531F,arnold:2013arXiv1310.2607A}. 
Thus, the outer halos cannot be studied by observing the stellar light directly. 

However, there are several reasons why the outer regions are of interest, 
especially as this is where the dark matter becomes dominant and its properties can be tested. 
\citet{romanowsky:2003Sci...301.1696R} for example found a steep Keplerian decline
in the velocity dispersion of NGC 3379, resulting in
discussions about the existence of a canonical dark halo \citep{dekel:2005Natur.437..707D,douglas:2007ApJ...664..257D}.
In addition, the kinematics of the stellar component in the regions far away from the center 
are most likely to preserve some indications of the formation history of the galaxy. 
For example, in S0-galaxies, the distinction between random motion and the ordered motion of 
a kinematically cold stellar disk component 
can reveal if the galaxy is actually a 
fading spiral galaxy or was formed from a minor merger, as discussed by \citet{noordermeer:2008MNRAS.384..943N,cortesi:2013MNRAS.432.1010C}. 

In order to study ETGs to larger radii, tracers are needed. 
These tracers need to emit enough light 
to be detectable as single objects even at large distances. 
Currently the most important tracers are planetary nebulae (PNe) and globular clusters (GCs). 
Planetary nebulae emit a large amount of their light in the [\ion{O}{3}]$\lambda 5007$ line and 
can therefore be observed out to many effective radii, as seen in the famous example of NGC 5128 
where planetary nebulae have been found
out to 20 kpc by \citet{hui:1995ApJ...449..592H} and out to 80 kpc by \citet{peng:2004ApJ...602..685P}. 
With spectrographs it is now possible to obtain not only the positions but also the line-of-sight velocities of 
hundreds of objects far away from the center (e.g. \citet{mendez:2009ApJ...691..228M}: 591 PNe in NGC 4697,  
\citet{coccato:2009MNRAS.394.1249C}: 450 PNe in NGC 4374). 

Another possible tracer for the stellar dynamics in the outskirts of ETGs is GCs. 
Two groups of GCs can be distinguished: red, metal-rich GCs and blue, metal-poor GCs.
Studies by \citet{schuberth:2010A&A...513A..52S} and others provide evidence that 
the stellar field population is traced by red GCs, whereas blue globular clusters 
may have been accreted later. 
A total of more than 2500 GCs have been studied recently 
by \citet{pota:2013MNRAS.428..389P} in 12 nearby ETGs. 

In numerical simulations, one is not restricted by the decreasing surface brightness, but by resolution effects. 
However, with improved numerical techniques, the analysis of small scale details of ETGs is now possible. 
For example, the shell-structure of ETGs, 
first observed in NGC 1316 by \citet{malin:1977AASPB..16...10M}, 
could be understood 
to be due to disruption of an accreted galaxy by mergers (e.g. 
\citealt{toomre:1978IAUS...79..109T,schweizer:1986Sci...231..227S,hernquist:1988ApJ...331..682H,binneytremaine,cooper:2011ApJ...743L..21C})
into the host galaxy.  
Therefore, these shell-structures can reveal information about the merger history of the galaxy.

To have a further indicator for the formation history of ETGs, we study in this Letter the velocity dispersions of
merger remnants. The goal is to find residual kinematic signatures of the progenitor galaxies that are still detectable despite the violent relaxation
experienced by the merging galaxies.

%2.
\section{Simulations}
We study ten ETGs formed in isolated major mergers of both 
spiral-spiral and spiral-elliptical galaxies. 
For the simulations, we used the parallel TreeSPH-code Gadget-2 \citep{springel:2005MNRAS.364..1105S},
in which energy and entropy are manifestly conserved, including radiative cooling of a 
primordial hydrogen-helium composition. 
Star formation and supernova feedback were included, using
the self-regulated model of \citet{springel:2003MNRAS.339..289S}. 
The description of the interstellar medium is based on a two-component model, 
where cool clouds are embedded in a surrounding medium of hot gas \citep{mcKee:1977ApJ...218..148M,johansson:2006MNRAS.371.1519J}.

Nine out of our ten galaxies include a black hole. For modelling its feedback, we used the model of 
\citet{springel:2005MNRAS.361..776S}. 
To guarantee efficient merging in the simulations, BHs merge instantaneously as soon as one BH is within 
the other BH's smoothing length and its velocity has become smaller than the local 
sound speed of the surrounding particles. 
The progenitor disk galaxies are embedded in Hernquist like dark matter halos with concentration parameters $c_s=9$ of the 
corresponding NFW-halo \citep{navarro:1997ApJ...490..493N}. The baryonic disk scale-radius is $r_{D}=3.5 \ \rm kpc$.
For more details on the simulations, see  \citet{johansson:2009ApJ...707L.184J},  \citet{johansson:2009ApJ...690..802J} and \citet{remus:2013ApJ...766...71R}.

Our sample consists of 5 spiral-spiral mergers with a mass ratio of 1:1 for the progenitors 
and 4 mergers with a ratio of 3:1, as well as one mixed merger (spiral and elliptical formed by a 3:1 spiral-spiral merger). 
The merger parameters are described in detail in Table \ref{tab:setup}.
To demonstrate our analysis, we use the 1:1 spiral-spiral merger 11 OBH 13 as 
an example galaxy, which has the following initial setup: 
inclinations $i_1 = -109^{\circ}$ and $i_2 = 180^{\circ}$, pericenter arguments $\omega_1 = 60^{\circ}$ and $\omega_2 = 0^{\circ}$, $v_{vir} = 160 \ \rm km~s^{-1}$.
\begin{table*}
\caption{Binary merger simulation sample at a timestep of 3~Gyr}             
\label{tab:setup}      
\centering          
\begin{tabular}{l c c c c c c c c c c c}
\hline\hline       
                      
Model & Ratio\tablenotemark{(a)} & Orbit\tablenotemark{(b)} & $f_{\mathrm{gas}}$\tablenotemark{(c)}  &  bulge \tablenotemark{(d)}&  BH \tablenotemark{(e)} & $v_{\mathrm{vir}}$\tablenotemark{(f)} & $\Reff$\tablenotemark{(g)} & $\beta$ \tablenotemark{(h)} \\

\hline
 11 NB NG 13    & 1:1 &  G13 & 0.0 & no & yes & 160 & 8.48 &	-0.17	 \\ 
 11 NB OBH 13    & 1:1 &  G13 & 0.2 & no & yes & 160 & 6.15 & 	-0.19	 \\ 
 11 NG 13    & 1:1 &  G13 & 0.0 & yes & yes &160 & 6.76 &  	-0.17	 \\
 11 OBH 09  & 1:1 &  G09 & 0.2 & yes & yes & 160 & 5.02 &  	-0.22	\\ 
 11 OBH 13 & 1:1 & G13 & 0.2 & yes & yes & 160 & 5.20 &  	-0.20	 \\
 31 ASF 01     & 3:1 & G01 & 0.2 & yes & no & 160 & 4.59  &  	-0.21	 \\ 
 31 O8BH 13     & 3:1 & G13 & 0.8 & yes & yes & 160 & 2.56 &  	-0.19	 \\
 31 OBH 09 320     & 3:1 & G09 & 0.2 & yes & yes & 320 & 10.42 & -0.10 	 \\ 
 31 OBH 13	& 3:1 & G13 & 0.2 & yes & yes & 160 & 5.22 &  	-0.18	 \\
 mix 11 OBH 13     & 1:1 & G13 & 0.2 & yes & yes & 160 & 6.32 & -0.18	 \\ 

\hline                  
\end{tabular}
\tablecomments{(a) initial mass ratio of the two galaxies;
(b) Orbit type according to \citet{naab:2003ApJ...597..893N,naab:2006ApJ...636L..81N,khochfar:2006A&A...445..403K};
(c) Initial gas fraction of the disks of the progenitor galaxies;
(d) Do the progenitor galaxies contain a Bulge?
(e) Do the progenitor galaxies contain Black Holes?
(f) Initial virial velocity in km $s^{-1}$;
(g) Effective Radius in kpc at 3 Gyrs;
(h) Slope for fitting $\sigma \propto r^\beta$ to the stellar component of the velocity dispersion at 3 Gyrs in the range of 0--50 kpc;
}

\end{table*}

%3.
\section{Results}
Observations have shown that the projected velocity dispersion ( = root mean squared velocity) of the stellar component 
can in general be well-fitted by a power-law \citep{douglas:2007ApJ...664..257D,napolitano:2009MNRAS.393..329N} 
\begin{equation}
\sigma \propto r^{\beta},
\end{equation}
\begin{figure}
\begin{center}
  \includegraphics[width=1\columnwidth]{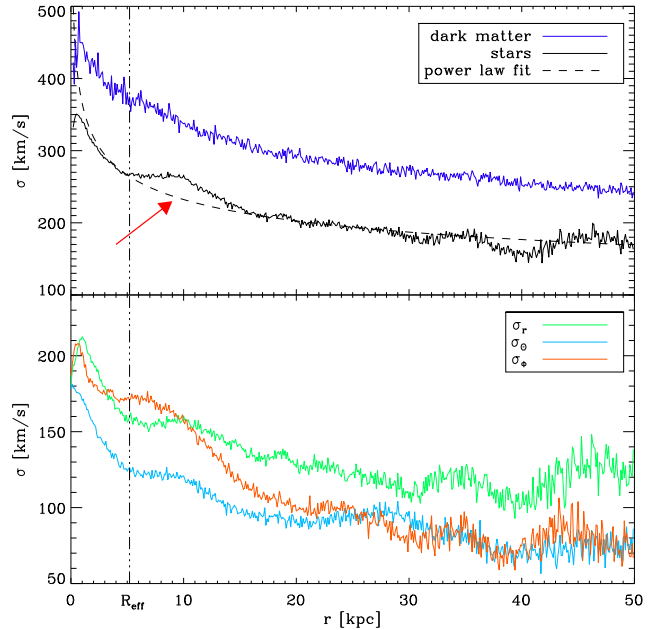}
  \caption{Upper panel: Stellar (black) and dark matter (blue) intrinsic velocity dispersion 
as function of radius for galaxy 11 OBH 13. 
For the stellar component, the power law fit is included (black dashed curve). 
The $\sigma$-bump (red arrow) as a positive deviation from a power-law behavior only shows in the stellar component.
Lower panel: Radial (green), tangential (blue) and azimuthal (orange) components 
of the stellar velocity dispersion.
}
  {\label{fig:sigma}}
\end{center}
\end{figure}

We calculate the intrinsic velocity dispersion profiles for our 
spheroidals and fit a power-law to the stellar component. 
Therefore, each directional component of the velocity dispersion is computed in 
radial bins as  
\begin{equation}
\sigma_i (r) = \sqrt{ \frac{\sum v_i(r)^2  }{N} - 
\left( \frac{  \sum v_i(r) }{N}    \right)^2 },
\label{eq:sigma}
\end{equation}
where the sum runs over all particles within the bin.
The intrinsic velocity dispersion is then calculated as   
\begin{equation}
\sigma = \sqrt{\sigma_1^2 + \sigma_2^2 + \sigma_3^2}.
\end{equation}
The upper panel of Figure~\ref{fig:sigma} shows the velocity dispersion of 
the stellar and dark matter component of 
our example galaxy, 11 OBH 13, as a function of radius, 
together with a power-law fit to the stellar component. 
For our sample of spheroidals, we find the mean of the power law exponent to be $\beta = - 0.18 \pm 0.04$  for the stellar component (see Table \ref{tab:setup}). This is in agreement with the stellar slope of \citet{dekel:2005Natur.437..707D} ($\beta = -0.4 \pm 0.1$ in projection) and slightly below the values found in hydrodynamical cosmological zoom--in simulations ($\beta = -0.05 \pm 0.06$) and large--scale hydrodynamical cosmological simulations ($\beta = -0.003 \pm 0.135$) of \citet{remus:2013ApJ...766...71R}.

We include more than 680 000 (450 000) stellar particles in each 1:1 (3:1) 
ETG, with each bin having a binwidth of 0.1 kpc and containing 
more than 1000 particles for radii up to 20 kpc.
To ensure that our results do not depend on the binning, 
we tested equal--mass bins (1500 particles per bin) and other 
equal--radius binwidths, 
but found no differences. 

\begin{figure*}
\begin{center}
  \includegraphics[width=2\columnwidth]{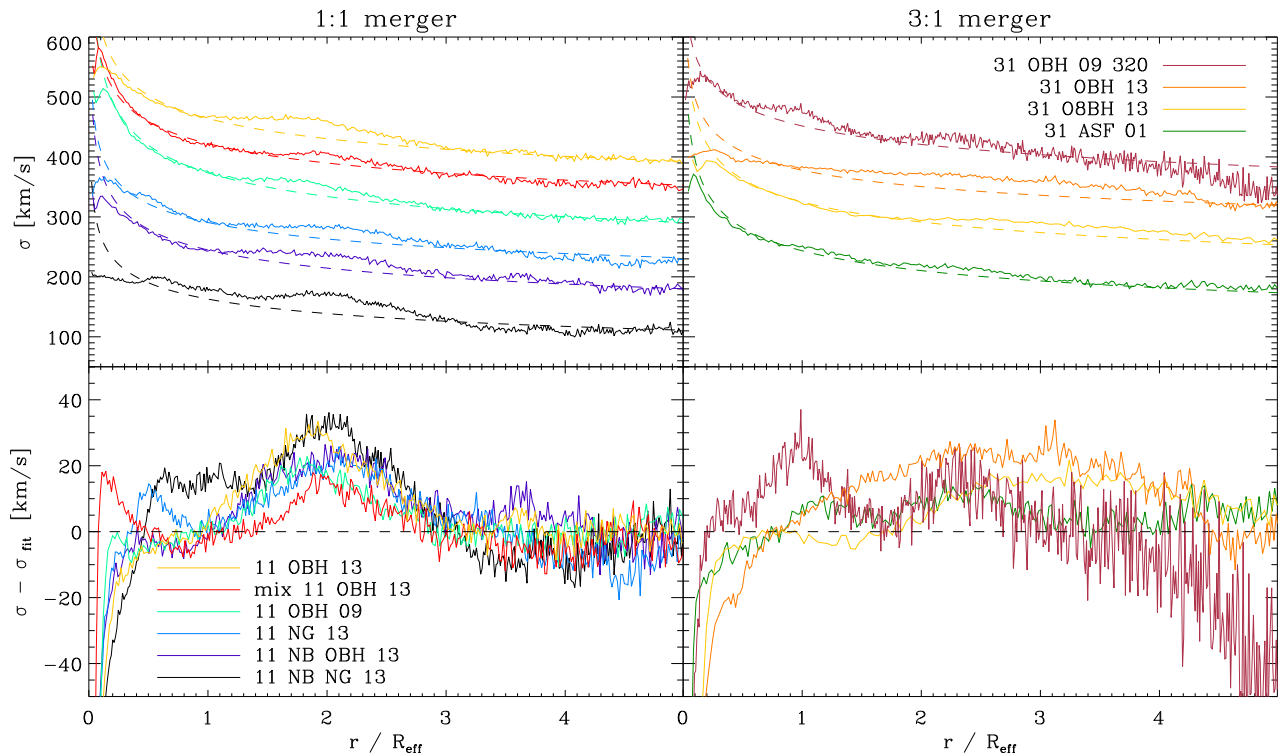} 
  \caption{Left column: 1:1 mergers, right column: 3:1 mergers (all at 3 Gyrs). 
Upper panels: Velocity dispersion of the different spheroids (solid lines) and their corresponding power laws (dashed lines) as function of effective radius. 
For better readability, we shift the dispersion profiles for all mergers except 11 NB OBH 13, 31 ASF 01 and 31 O8BH 09 320 by 
factors of 50. 
The lower panel displays the difference between the velocity dispersion and its best fit power-law. 
}
  {\label{fig:3x3}}
\end{center}
\end{figure*}

As can be seen in Figure~\ref{fig:sigma}, the stellar velocity dispersion generally follows the power law, 
but shows some deviations. 
The innermost deviation of the velocity dispersion from a power-law within 2 kpc
is due to very bound bulge-stars 
in the deep gravitational potential at the galaxies' center. 
Here, we are not interested in the bulge component and therefore neglect the innermost radii.
Other deviations from the power-law are caused by shell-structures. 
Shells can form in major mergers, e.g. described by \citet{cooper:2011ApJ...743L..21C},
leading to an ejection of stars 
in density waves \citep{schweizer:1986Sci...231..227S}. 
Therefore, the velocity dispersion varies at the radii where shells are present. 

We distinguish between the deviation in the range of 5 to 15 kpc and the 
oscillations at larger radii. The latter are clearly linked to 
regions of high particle density, 
visually identified as shell-structures. 
In the region between 5 and 15 kpc, where $\sigma$ stays constant up to 11 kpc  
and then decreases rapidly to follow the power law again, no shell feature can be identified in the simulations. 
Therefore, the deviation must have another origin and we call it $\sigma$-bump. 

\begin{figure*}
\begin{center}
  \includegraphics[width=2\columnwidth]{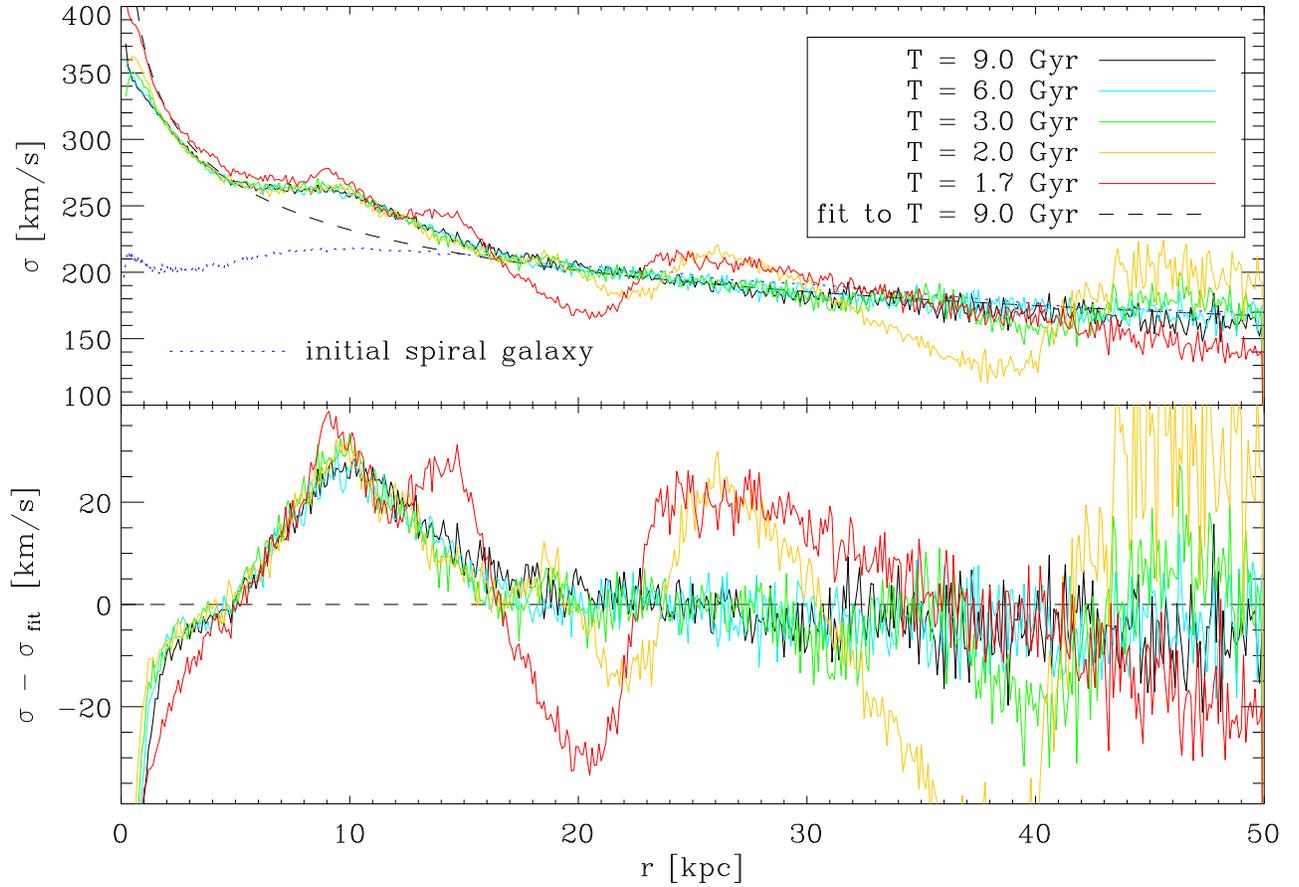}
  \caption{ Upper panel: Intrinsic stellar velocity dispersion of the example galaxy against radius, 
different colors indicate different time steps of the simulation 
(merger encounter at 1.5 Gyrs),
intrinsic stellar velocity dispersion of the progenitor spiral galaxy (blue dashed line).
Lower panel: Difference of the intrinsic stellar velocity dispersion to the power-law fit of the last snapshot 
for different time steps (colors same as upper panel). 
}
  {\label{fig:time}}
\end{center}
\end{figure*}

In the dark matter component, the velocity dispersion is much higher,
following a power law without significant deviations,
showing no signature of the $\sigma$-bump (see upper panel of Figure~\ref{fig:sigma}).
The stellar components of our other nine galaxies also 
show a $\sigma$-bump (see Figure~\ref{fig:3x3}), 
including some simulations without gas or without a bulge.
The mergers with mass ratio of 1:1 show a more prominent $\sigma$-bump than the mergers with mass ratio of 3:1. 
We therefore conclude that the $\sigma$-bump is a 
purely stellar feature common in 
our major merger sample and is dependent on the mass ratio of the progenitor galaxies. 

The lower panel of Figure~\ref{fig:sigma} shows the radial, tangential 
and azimuthal component of the velocity dispersion of our example galaxy against radius.
The $\sigma$-bump is most prominent in the azimuthal component, whereas for larger radii, 
the deviations from the power law are dominated by the radial component and  
therefore can be associated with the drift of the shell-structures. 
A feature like the $\sigma$-bump, which is most prominent in the azimuthal component, 
could be caused by a disk-structure embedded in the ETG. 

In order to understand the origin of the $\sigma$-bump and to analyse its shape and size, 
we investigate the velocity dispersion at different timesteps. 
The majority of our simulations run for 3.0 Gyrs, 
with the merger taking place at about 1.5 Gyrs for all galaxies, simulation 11 OBH 13 is run for 9.0 Gyrs.

The upper panel of Figure~\ref{fig:time} shows the stellar velocity dispersion of 
galaxy 11 OBH 13 at different times, indicated by different colors, 
from 1.7 Gyrs (red curve) to 9.0 Gyrs (black curve). 
The velocity dispersion of the progenitor spiral galaxy is included as
blue dotted line. The $\sigma$-bump is not present in the progenitor disks, it is a feature of the merger remnant alone. 
At the first shown timestep, it is still forming, but from 2.0 Gyrs onwards, the
$\sigma$-bump remains constant at all times, 
and thus differs clearly from all the other deviations present at larger radii, which vary or propagate outwards 
and disappear after some time. 
This is emphasized in the lower panel of Figure~\ref{fig:time}, where the 
difference of the stellar velocity dispersion with respect to the power law fit of the final timestep is shown. 
Here, we can see more clearly that the $\sigma$-bump remains the same in size and shape, 
while the shell-structures vanish with time. 

%4
\section{Comparison to Observations}
\begin{figure*}
\begin{center}
  \includegraphics[width=2\columnwidth]{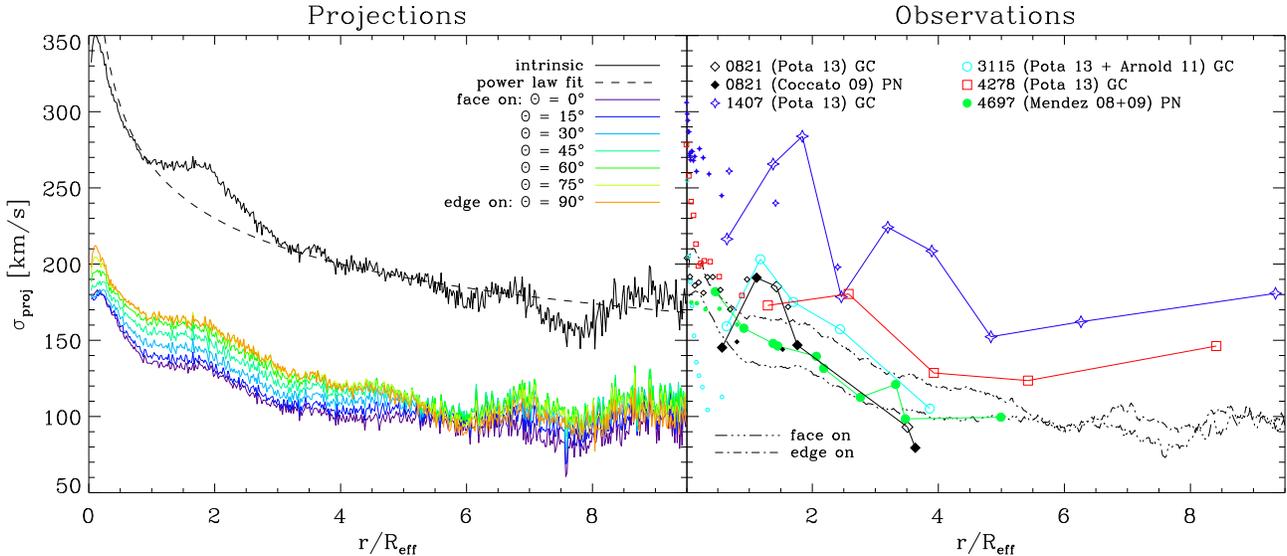}
  \caption{Left panel: Stellar velocity dispersion against radius of example galaxy 11 OBH 13. 
Black line: intrinsic stellar component, black dashed line: power law fit to intrinsic stellar component, 
dark violet: face on projection, orange: edge on projection, other colors: 
projections at different angles. 
\\Right panel: GCs and PNe are shown by large symbols, additional stellar data by small symbols of the same kind and color (stellar data for NGC 821 from \citet{forestell:2010ApJ...716..370F} (open diamonds) and 
\citet{proctor:2009MNRAS.398...91P} (filled diamonds), for NGC 1407 from \citet{proctor:2009MNRAS.398...91P} (open stars) 
and \citet{spolaor:2008MNRAS.385..667S} (filled stars), for NGC 3115 from \citet{norris:2006MNRAS.367..815N}, 
for NGC 4278 from \citet{marel:1993ApJ...407..525V}, for NGC 4697 from \citet{binney:1990ApJ...361...78B}). We include the face--on and edge--on projections of the simulated 
example galaxy as dash--dotted lines. 
}
  {\label{fig:proj}}
\end{center}
\end{figure*}

To compare our results to observations 
it is not sufficient to just consider the intrinsic stellar velocity dispersion, but also its projections. 
Figure~\ref{fig:proj} shows the line-of-sight velocity dispersion 
for different projections from face-on ($0^\circ$) to edge-on ($90^\circ$) in different colors against radius
for our example galaxy, 
as well as the intrinsic velocity dispersion and its power-law fit. 
The $\sigma$-bump can be seen in all line-of-sight velocity projections, although with a somewhat lower amplitude and 
should therefore be detectable by observations. 

We compare our simulations with results from radial velocity measurements of PNe and red GCs in ETGs, 
as the red GCs are presumed to trace the stellar component of ETGs. 
We use the position and velocity measurements of the observers to calculate the velocity dispersion as a function of radial distance. 
For most of the observed galaxies currently available, the observational data sets are not sufficient: 
either the total sample of tracers contains less than 200 objects, 
or the data mainly trace radii at which we do not expect to see the $\sigma$-bump.

\citet{pota:2013MNRAS.428..389P} recently published the kinematics of a sample of globular clusters in 12 ETGs 
from a spectroscopic survey (see also \citealt{strader:2011ApJS..197...33S,arnold:2011ApJ...736L..26A,foster:2011MNRAS.415.3393F}). 
The sample of PNe data from \citet{coccato:2009MNRAS.394.1249C} is of similar size, however here we focus on the sample of \citet{pota:2013MNRAS.428..389P}. 
We find a $\sigma$-bump in four of the twelve galaxies (NGC 821, NGC 1407, NGC 3115, NGC 4278) from this sample, 
while  three other galaxies
do not show a significant comparable feature 
but a constant or decreasing velocity dispersion. 
For the remaining five galaxies we cannot draw any firm conclusion
as we are limited by low-number statistics 
(37 and 42 red GCs in NGCs 1400 and 2768, respectively, and 21 GCs in NGC 7457), 
the feature varies a lot with binsize (NGC 4486) or is observed at too large radii (NGC 5846).
We include in Figure~\ref{fig:proj} the observed galaxies which show a $\sigma$-bump 
at the location of the $\sigma$-bump of our simulated galaxy 11 OBH 13: 

For NGC 821, both PNe and red GCs trace 
a $\sigma$-bump behavior between 0.5 and 1.8 $\Reff$ ($\Reff= 5.79$ kpc, 
PNe data from \citealt{coccato:2009MNRAS.394.1249C}, 
GCs data from \citealt{pota:2013MNRAS.428..389P}, stellar data 
from \citealt{proctor:2009MNRAS.398...91P,forestell:2010ApJ...716..370F}). 
It is an isolated E6 galaxy, with a velocity dispersion 
that generally shows a rapid decrease with radius \citep{romanowsky:2003Sci...301.1696R} 
and kinematic and photometric signatures of an edge-on stellar disk \citep{proctor:2009MNRAS.398...91P}. 

NGC 1407 is a  large, round, massive elliptical with a large velocity dispersion 
\citep{goudfrooij:1994A&AS..104..179G}. 
Following \citet{pota:2013MNRAS.428..389P}, we choose the red GCs out of 
356 confirmed GCs and bin them in equal-number bins of about 20 objects. 
The velocity dispersion shows two positive deviations - one in the interval of 
0.6 $\Reff$ and 2.5 $\Reff$, the other one from 2.5 $\Reff$ to 4.8 $\Reff$ ($\Reff= 9.36$ kpc, 
GCs data from 
\citealt{pota:2013MNRAS.428..389P}, stellar data from \citealt{proctor:2009MNRAS.398...91P,spolaor:2008MNRAS.385..667S}). 

The GCs data of NGC 3115 were first presented by 
\citet{arnold:2011ApJ...736L..26A}. This S0-galaxy contains a chemically enriched 
and kinematically distinct stellar disk \citep{norris:2006MNRAS.367..815N}.
The $\sigma$-bump
can be seen in the range of 0.7 to 2.2 $\Reff$ ($\Reff$=3.87 kpc).
For NGC 4278 (GCs data from \citealt{pota:2013MNRAS.428..389P},
stellar data from \citealt{marel:1993ApJ...407..525V}) 
we observe a $\sigma$-bump in a large range: from 
1.5 to 3.8 $\Reff$ ($\Reff$ = 2.58 kpc). This galaxy has a large HI disk
and a dusty patch \citep{goudfrooij:1994A&AS..104..179G}.

Additionally, we found a $\sigma$-bump feature in the galaxy NGC 4697. 
For this galaxy we found the most extended data set, with 218 PNe from \citet{mendez:2009ApJ...691..228M} and 
531 PNe from \citet{mendez:2001ApJ...563..135M,mendez:2008ApJS..175..522M}. 
The kinematic information covers the central area as well as the outer regions of the galaxy, 
showing a $\sigma$-bump behavior between 0.9 and 2.7 $\Reff$ ($\Reff= 3.36$ kpc,
stellar data from \citealt{binney:1990ApJ...361...78B}).
NGC 4697 is an E4-5 galaxy with a stellar disk along the major axis 
\citep{carter:1987ApJ...312..514C,goudfrooij:1994A&AS..104..179G} 
and at least two PNe subpopulations \citep{sambhus:2006AJ....131..837S}.

We thus conclude that the $\sigma$-bump is a feature that can be observed 
with tracers in the outer parts of ETGs in the stellar component, 
however a statistically significant amount of tracers is needed. 

%5. 
\section{Summary and Discussion}
We have identified a new feature in the kinematics of ETGs, 
which can be seen in all spheroidals resulting from our sample of ten simulated isolated major mergers. 
The azimuthal component of the velocity dispersion 
contributes the most to the $\sigma$-bump, whereas shells are dominated by the 
radial dispersion component. 
This $\sigma$-bump can already be seen shortly after the merging event and remains 
stable with time, while other features such as shells vanish after a few Gyrs. 
We found the $\sigma$-bump to be a purely stellar feature which is not mirrored by 
the velocity dispersion of the dark matter component. The $\sigma$-bump is most prominent in 1:1 mergers and therefore might be a signature for major mergers.

Observations of some ETGs such as NGC 821, NGC 1407, NGC 3115, NGC 4278 and NGC 4697 show 
a positive deviation of the velocity dispersion in the same $\Reff$-range as our galaxies. 
In three out of these five galaxies, a stellar disc is observed, 
while a fourth galaxy has a HI disk. All galaxies from \citet{pota:2013MNRAS.428..389P} that show a 
 $\sigma$-bump are fast rotators as well as NGC 4697 \citep{emsellem:2011MNRAS.414..888E}. 
In the sample of \citet{pota:2013MNRAS.428..389P}, two of the three galaxies that do not show $\sigma$-bump are also 
fast rotators, one is a slow rotator. 

The fact that the $\sigma$-bump is also present in galaxies which have been simulated without  gas 
or  bulge components suggests that it is a remnant of the kinematics of the disk stars of the progenitor galaxy. 
The region of the $\sigma$-bump is interesting, as it corresponds to the size of the disk of the progenitor galaxy.
Between 7 kpc and 14 kpc (23 kpc for 31 OBH 09 320), the dark matter begins to dominate 
over the stellar component, which might also influence the dynamics of the stellar component.  
The reasons for the presence or absence of the $\sigma$-bump 
thus remain to be investigated. 

In a future study (Schauer et al., in prep) we will investigate a larger sample of simulated galaxies, including spheroidals from minor mergers and 
cosmological simulations, in order to survey if the $\sigma$-bump is present 
only in major mergers and how it relates to the disk of the progenitor galaxies. 
It is interesting that we see a $\sigma$-bump also in the Sbc dry merger with 0\% gas by \citet{dekel:2005Natur.437..707D}, but in none of their other mergers, and at least in one of the galaxies with mass ratio of 1:1 from \citet{jesseit:2007MNRAS.376..997J}.

The velocity dispersion is a quantity for which the accuracy strongly depends on the number of observed tracer objects.  
To understand which signatures about the formation history and evolution of ETGs are retained by their outer halos, more detailed observations of the outskirts of ETGs are required, 
especially larger observational tracer samples. 

\begin{acknowledgements}
We thank the referee for many helpful comments. 
R-S.R. acknowledges a grant from the International Max-Planck Research School of Astrophysics (IMPRS).
A.T.P.S., R-S.R. and A.B. acknowledge financial support from the cluster of excellence 
``Origin and Structure of the Universe''. 
P.H.J. acknowledges the support of the Research Funds of the University of Helsinki.
\end{acknowledgements}

\end{document}